\documentstyle[eqsecnum,aps]{revtex}

\input epsf
\epsfverbosetrue
\begin{document}
\begin{titlepage}
\begin{flushright}
HIP -- 1998 -- 01 / TH\\
\today
\end{flushright}

\begin{centering}
\vfill
{\Large\bf An interquark potential  model for multi-quark systems}
\vspace{1cm}

A. M. Green\footnote[1]{E-mail: {\tt anthony.green@helsinki.fi}}
and
P. Pennanen\footnote[2]{E-mail: {\tt petrus@hip.fi}}

\vspace{0.25cm}

{\em Department of Physics and Helsinki Institute of Physics \\
P.O. Box 9 \\ FIN-00014 University of Helsinki \\ Finland }
\vspace{1.5cm}

{\bf Abstract}\\
A potential model for four interacting quarks is constructed in SU(2)
from six basis states -- the three partitions into quark pairs, where
the gluon field is either in its ground state or first excited
state. With four independent parameters to describe the interactions
connecting these basis states, it is possible to fit 100 pieces of 
data -- the ground and first excited states of configurations from six
different four-quark geometries calculated on a $16^3\times 32$ lattice. 

\vspace{0.5cm}
\end{centering}

\noindent

\vfill

\end{titlepage}

\section{Introduction}

Over the years there has been much progress in lattice QCD. However,
this has been restricted to systems with only a few quarks -- upto three
in most cases. On the other hand, in particle and nuclear
physics there is also considerable interest in few- and multi-{\em hadron}
systems beginning with the possibility of bound $K\bar{K}$ states. 
It seems unlikely that direct results for any but the simplest of these 
hadron systems could 
be obtained in the near future from lattice QCD simulations. Our approach is 
thus to
construct a model that reproduces  lattice results for the simplest 
multi-quark systems in a way that can be relatively easily extended to more
complex cases. The lattice data has been obtained earlier by the Helsinki 
group \cite{glpm:96}(and references therein) as energies of four static 
quarks in various geometries, such as quarks at the corners of squares, 
rectangles, tetrahedra and some other less symmetric geometries. 
This 'experimental' data is then to be explained by a model. 
The choice of geometries is hopefully general enough for the model to 
reproduce also the energies for geometries intermediate to those actually 
considered.

It should be pointed out that the 'experimental' data is not for full
QCD. For numerical reasons, the SU(2) gauge group is used instead of SU(3), 
the use of the latter requiring an order of magnitude
increase in computer resources. However, the indications are that, for
the present type of discussion, SU(2) suffices. Another approximation
is taking the quarks to have an infinite mass. This means that our
constituent quarks are static and that there are no sea quarks (the so-called
quenched approximation). 
At present, this limit is now being partially removed by applying 
the techniques of 
Ref.~\cite{mic:97} to a system of two light and two heavy (static) quarks. 
The sea quarks will appear through using gauge configurations 
generated by the UKQCD collaboration for full QCD. However, in this
paper, we only 
attempt to understand systems with four static SU(2) quarks in the quenched 
approximation.

Any model that can be extended to multi-quark systems must presumably
depend explicitly only on the quark degrees of freedom, with the gluon
degrees of freedom entering only implicitly. This is the same philosophy
that has been successful for interacting multi-nucleon sytems. In the
latter, the meson fields generating the interactions are 'summarized' as
internucleon potentials, which take the form of two-nucleon potentials
and, to a lesser extent, three- and four-nucleon potentials. However, 
an important difference between multi-quark and multi-nucleon systems
is the strongly self-interacting nature of the gluon fields mediating 
the interaction in the former case. It is, therefore, not at all clear that 
{\em any} effective model defined in terms of only quark degrees of
freedom will be successful. It is the purpose of this article to see to
what extent a model can be developed  for the four-quark case. If
this attempt fails, then there is no point in expecting it to work in even
more complex quark problems; a successful model for four quarks is
necessary but not sufficient before considering any extension to even
more quarks.

It should be added that, in spite of the problems outlined above, many
groups still consider that multi-quark systems can be treated using
simply the basic two-quark potential e.g.~\cite{pep:96,sak:97}. In the opinion 
of the authors -- and backed up by their earlier works -- there seems to be
no justification for such an approach.

In Section 2 three versions of the model, starting with the simplest, 
are introduced and in Section 3 results are given. 

\section{ The Model with 2,  3 or 6 basis states}

Since the four quarks in the lattice calculation are static, the
corresponding model
should not contain any kinetic energy. Also, because we only consider SU(2),
there is no distinction between the group properties of quarks($q$) and
antiquarks($\bar{q}$).  Four such quarks can then be partitioned as pairs
in three different ways
\begin{equation}
\label{ABC}
A=(q_1q_3)(q_2q_4), \ \ B=(q_1q_4)(q_2q_3) \ \ {\rm and} \ \ C=(q_1q_2)(q_3q_4),
\end{equation}
where each $(q_iq_j)$ is a colour singlet. However, these three basis
states are not orthogonal to each other. Also, remembering the fact that
the quarks are indeed fermions gives, in the weak coupling limit, the
condition in the appendix of Ref.~\cite{gms:94} 
\begin{equation}
\label{A+B+C}
|A+B+C>=0.
\end{equation}
Since $<A|A>=<B|B>=<C|C>=1$, we get -- in this limit -- the equalities \\
$<A|B>=<B|C>=<A|C>=-1/2$.

\subsection{2 basis states}
\label{2.1}
Restricting the basis to only the two states $A$ and $B$
 leads to the  normalisation matrix 
\begin{equation}
\label{N}
{\bf N}=\left(\begin{array}{ll}
1&-1/2\\
-1/2&1\end{array}\right).
\end{equation}

In addition to this there is the potential matrix
\begin{equation}
\label{V}
{\bf V}=\left(\begin{array}{cc}
v_{AA} & V_{AB}\\
V_{BA}& v_{BB} \end{array}\right).
\end{equation}
In the weak coupling limit -- when all quarks are close to each other --
we expect the potential matrix to simply become that form appropriate
for one-gluon exchange  i.e.
\begin{equation}
\label{OGE}
V_{ij}=-\frac{1}{3}\sum_{i\le j} \vec{\tau_i} \cdot \vec{\tau_j} v_{ij},
\end{equation}
where $v_{ij}=-e/r_{ij}$. Then $V_{AA}=v_{13}+v_{24}$ etc. and
\begin{equation}
\label{VAB}
V_{AB}=V_{BA}=
-\frac{1}{2}\left(v_{13} +v_{24} +v_{14}+v_{23} - v_{12}-v_{34} \right).
\end{equation}
The object is then to compare the eigenvalues of 
\begin{equation}
\label{VN}
\left[{\bf V}-E(4) {\bf N}\right]=0
\end{equation}  
 with the lattice
results -- the success or failure of the model being to what extent the
two agree. Here $E(4)$ is the total four-quark energy. However, in
practice, it is more convenient to deal with the four-quark {\em binding}
energy $E$ defined as $E=E(4)-V_{AA}$, assuming that the $A$ partition has
the lowest energy. A perturbative calculation to $O(\alpha^2)$ has 
reproduced this two-state model \cite{lan:95}.

If all three basis states are included, the model has a problem since the
matrix in Eq.~\ref{VN} is singular for the obvious reason that
$|A+B+C>=0$.
In some of our earlier work this was interpreted to mean that it was unnecessary
to include all three states and so the symmetry was broken by keeping
the two states with the lowest energy, let us say , $A$ and $B$.
A similar thing also occurred in the lattice simulations. There it was
found that the energy of the lowest state was always the same in both a
2$\times$2 and 3$\times$3 description, providing $A$ or $B$ had the lowest
energy. In addition the energy of the second state was, in most cases, more or less the same -- the largest
difference occurring with the tetrahedral geometry. Another modification
to the basic model in Eqs.~\ref{OGE}--\ref{VN} is to replace the original
one-gluon-exchange potential by the full two-quark potential
\begin{equation}
\label{V2}
v_{ij}=-e/r_{ij}+b_sr_{ij}+c.
\end{equation}
This is now a move away from the extreme weak coupling limit. In
principle, everything is now known, since the parameters $e,b_s,c$ in
$v_{ij}$ can be determined from the same lattice calculation as the
four-quark energies. However, this model fails badly, since it still
contains the van der Waals long range forces. The main feature of this
failure is that the lowest eigenvalue ($E_0$) is too low and the
other ($E_1$)  too
high. This suggests that the off-diagonal potential $V_{AB}$ is too
large. However, as we go away from the weak coupling limit, it is
expected that $<A|B>$
will decrease until eventually, in the extreme strong coupling limit, it
would vanish. This, therefore,  suggests that for some realistic
intermediate coupling the off-diagonal normalisation matrix element
$N_{AB}$ should also be reduced. This effect we simulate by introducing
a factor $f$, which decreases as the distance between the quarks
increases i.e.
\begin{equation}
\label{Nf}
<A|B>=-f/2.
\end{equation}
For internal consistency, this same factor must also be introduced 
into  $V_{AB}$ i.e.
\begin{equation}
\label{Vf}
V_{AB}\rightarrow f V_{AB} 
\end{equation}
  otherwise the eigenvalues would depend on the self-energy term $c$
in Eq.~\ref{V2}. In Ref.~\cite{glpm:96} this model had the good feature that,
when fitting the data ($E_0,E_1$) for a given square, only a single $f$ was
necessary to fit both energies. Of course, $f$ was dependent on the size
of the square, but a reasonable  parametrization was
\begin{equation}
\label{f}
f(Ia)=\exp(-b_s k_f S),
\end{equation}
where $b_s$ was the string energy of Eq.~\ref{V2}, $S$ the area of the
square and $k_f\approx 0.5$. The form of this parametrization  was
motivated by strong coupling ideas \cite{mor:89,ale:90,gre:90}. 
The original hope was that, with $k_f$ determined from the squares, the model 
should automatically also fit other geometries with $S$ being the
appropriate area contained by the four quarks. When the four quarks lie
in a plane, the definition of $S$ is clear. However, in non-planar
cases the situation is more complicated and so here the area is simply taken
to be the average of the sum of the four triangular areas defined by  the
positions of the four quarks i.e. the faces of the tetrahedon. For
example, in the notation of Eq.~\ref{ABC}, the appropriate area $S(AB)$ for $f$ is
\begin{equation}
\label{Def:S}
S(AB)= 0.5[S(431)+S(432)+S(123)+S(124)],
\end{equation}
where $S(ijk)$ is the area of the triangle with corners at $i,j$ and
$k$. For planar geometries this simply reduces to the expected area. But
for non-planar cases this is only an approximation to $S(AB)$ -- a more
correct area being one that is not necessarily a combination of planar
areas  but of curved surfaces with 
minimum areas. These possibilities are discussed in Ref.~\cite{fur:96}. It
would be feasible to incorporate this refinement here, since only a few
$(\approx 50)$ such areas are needed. But for a general situation, in
which the positions of the quarks are integrated over, it would become
impractical, since the expression for the minimum area itself involves a double
 integration. In contrast, the area used here is an algebraic expression
and is, therefore, more  readily evaluated for any
geometry. The above model will be referred to as Version Ia.

This model has only one free parameter $k_f$ in $f$. Another possibility
with additional parameters $f_0,\ k_P$ is
\begin{equation}
\label{fIb}
f(Ib)=f_0\exp(-b_s k_f S+\sqrt{b_s} k_P P) \ \ {\rm (Version \ \ Ib)},
\end{equation}
where $P$ is the perimeter bounding $S$. 
However, as shown in Ref.~\cite{pen:96b},  this reduces in the continuum limit
 to the same as Version Ia -- the differences at $\beta =2.4$
being mainly due to lattice artefacts.   

Unfortunately, both of these models have the feature that, for regular
tetrahedra, they are unable to reproduce a degenerate ground state with a
{\em non-zero} energy, since  the two eigenvalues are
\begin{equation}
\label{E01}
E_0=-\frac{f/2}{1+f/2}[V_{CC}-V_{AA}] \ \ {\rm and} \ \
E_1=\frac{f/2}{1-f/2}[V_{CC}-V_{AA}] 
\end{equation}
and for regular tetrahedra $V_{CC}=V_{AA}$.

\subsection{3 basis states}
In the previous subsection only two of the possible three basis states
$A,B,C$ were used, the original reason for this being the condition in
Eq.~\ref{A+B+C}.
However, once the factor $f$ is introduced as in Eqs.~\ref{Nf}, \ref{Vf} this
condition no longer holds, so that all three states can be incorporated.
This apparently leads to the need for two more factors $f',f''$ defined by
\begin{equation}
\label{Nf'}
<A|C>=-f'/2 \ \ {\rm and} \ \ <B|C>=-f''/2.
\end{equation}
However, with the parametrization of $f$ as in Eq.~\ref{f} or \ref{fIb} and the
definition of $S$ as in Eq. \ref{Def:S}, it is seen that
$f'=f''=f$, since $S$ is simply proportional to the area of the faces of
the tetrahedron defined by the four quark positions and is {\em independent} of the
state combination used. Therefore, a $3\times 3$ model has the form where the
$N$ and $V$ matrices are:

\begin{equation}
\label{N33}
{\bf N}(f)=\left(\begin{array}{lll}
1&-f/2&-f/2\\
-f/2&1&-f/2\\
-f/2&-f/2&1\end{array}\right)
\ \ {\rm and} \ \ 
\end{equation}
\begin{equation}
{\bf V}(f)=\left(\begin{array}{ccc}
v_{AA} & fV_{AB}& fV_{AC}\\
fV_{BA}&v_{BB} & fV_{BC}\\
fV_{CA}&fV_{CB}&v_{CC} \end{array}\right).
\end{equation} 
Unfortunately, this model(II) also has some unpleasant features:

i) Again for regular tetrahedra all three eigenvalues are zero.

ii) For a linear geometry, since the 'appropriate' area as defined by 
Eq.~\ref{Def:S} vanishes, we get $f=1$ i.e. we are back to the weak coupling limit
and a singular matrix. 

iii) For squares the model gives $E_1=-E_0$, whereas the predictions of
Model I in Eqs.~\ref{E01} seem to be nearer the data.

The regular tetrahedron result is clear. There is only one energy
scale in the model, since all the $v_{ij}$ are the same. Therefore, there
can not be any excitations. The positive feature of this model is that
all three states are treated equally - a point that seems to be
necessary for regular tetrahedra.
\subsection{ 6 basis states}
\label{modII}
The above models both have trouble in describing regular tetrahedra.
This might be considered a minor point to worry about, since such a
configuration is very special. However, our philosophy is that, if {\em
any} configuration cannot be fitted, then the model fails, since then
there is no reason to  expect configurations not checked explicitly to
be fitted. An interesting feature of the regular tetrahedron data is
that the lowest state becomes {\em more} bound as the tetrahedron
increases in size with the magnitude of $E_0$ increasing from
  --0.0202(8) to --0.028(3) as the $d^3$ cube
containing the tetrahedron increases from $d=2$ to $d=4$. This is opposite
to what happens with squares, where the magnitude of $E_0$ decreases from
$-0.0572(4)$ to $-0.047(3)$ as $d$ increases from 2 to 5. This indicates that there is
coupling to some higher state(s) that becomes more effective as the size
increases and suggests that these higher states contain gluon
excitation with respect to the $A,B,C$ configurations. Model III,
introduced in Ref.~\cite{pen:96b},
therefore, extends Model II by adding three more states $A^*,B^*,C^*$,
where in analogy with Eq. \ref{ABC}
\begin{equation}
\label{A^*}
A^*=(q_1 q_3)_{E_u} (q_2 q_4)_{E_u} \ \ {\rm etc}.
\end{equation}
Here $(q_1 q_3 )_{E_u}$ denotes a state where the gluon field is excited
to the lowest state with the symmetry of the $E_u$ representation of the 
lattice symmetry group $D_{4h}$. Because it is an odd parity excitation, 
$A^*,B^*,C^*$ must contain two such states in order to have the
same parity as $A,B,C$.
The excitation energy of an $E_u$ state over its ground state($A_{1g}$)
counterpart is $\approx \pi/R$ for two quarks a distance $R$ apart.
As $R$ increases this excitation energy decreases making
the effect of the $A^*,B^*,C^*$ states more important, leading to the
effect mentioned above. Here we have
assumed that these states arise from a combination of excited states
with $E_u$. However, it is possible that they involve other excitations,
e.g.  
\begin{equation}
A^*=(q_1 q_3)_{A'_{1g}} (q_2 q_4) \ \ {\rm etc.},
\end{equation}
where the $A'_{1g}$ state is a gluonic excitation with the same quantum
numbers as the ground state ($A_{1g}$). For this case  the following
formalism would be essentially the same. Another possibility, which is not 
considered here, is that
the relevant excitations are flux configurations where all four quarks, instead
of two, are involved in  forming a colour singlet. 
In the strong coupling approximation such states would reduce to two-body
singlets due to Casimir scaling of the string tensions --
the string tension for a higher representation would be more than double
the value of the fundamental string tension, thus preventing junctions 
of two strings in the fundamental and one in the higher representation. This
would happen both in SU(2) and SU(3), the only exception being the unexcited 
$C$ state in SU(3), which would involve an antitriplet string.
 
There are now several  new matrix elements that need to be discussed for
${\bf N(f)}$ and ${\bf V(f)}$:

\vspace{0.3cm}

a) With  the addition of the $A^*,B^*,C^*$ states and 
the antisymmetry condition $A^*+B^*+C^*=0$ analogous to Eq.~\ref{ABC},
there are now two more functions $f^{a,c}$ defined as
\[<A^*|B^*>=<A^*|C^*>=<B^*|C^*>=-f^c/2  \ \ {\rm and} \ \ \]
\begin{equation}
<A^*|B>=<A^*|C>=.. \ \ {\rm etc.} \ \ .. =-f^a/2.
\end{equation}
Here it is assumed that $f^{a,c}$ are both dependent on $S$ as defined in
Eq.~\ref{Def:S}. Since $f^c$  involves only the excited states, it is
reasonable to expect it has a form similar to $f$ in Eq.~\ref{f} i.e.
\begin{equation}
\label{fc}
f^c=\exp(-b_sk_cS).
\end{equation}

\vspace{0.3cm}

b) By orthogonality $<A|A^*>=<B|B^*>=<C|C^*>=0$

\vspace{0.3cm}

c) In the weak coupling limit, from the $A^*+B^*+C^*=0$ condition,
we expect $<A|B^*>=<B|C^*>=.....=0$ at small distances.
To take this into account we parametrize $f^a$ as
\begin{equation}
\label{fa}
f^a=(f^a_1+b_s f^a_2 S)\exp(-b_sk_aS). 
\end{equation}
A partial justification of this is given in the Appendix and in Fig~6 
of Ref.~\cite{pen:96b}.
In the following fit all three parameters $f^a_1,f^a_2,k_a$ are
varied. However, it is found that $f^a_1$ is always 
consistent with zero -- as
expected from the above condition $<A|B^*>=....=0$. Therefore, in most of
this work  $f^a_1$ is fixed at zero.

\vspace{0.3cm}

d) For the potential matrix ${\bf V(f)}$ the diagonal matrix elements, 
after the lowest energy $V_{DD}$ amongst the
basis states is removed, are
\[<A^*|V-V_{DD}|A^*>=v^*(13)+v^*(24)-V_{DD}, \ \ {\rm etc.},\]
where  $v^*(ij)$ is the potential of the $E_u$ state -- a quantity also
measured on the lattice along with the four-quark energies and 

$V_{DD}={\rm min}[V_{AA}=v(13)+v(24), \ V_{BB}=v(14)+v(23), \ 
V_{CC}=v(12)+v(34)]$.

However, to
allow more freedom in the following fits, we introduce a parameter $b_0$,
where
\[<A^*|V-V_{DD}|A^*>= V_{AA}-V_{DD}+b_0 V^*_{AA} \ \ ,
 \ \ <B^*|V-V_{DD}|B^*>= V_{BB}-V_{DD}+b_0 V^*_{BB},\]
\begin{equation}
\label{C*C*}
<C^*|V-V_{DD}|C^*>=   V_{CC}-V_{DD}+b_0 V^*_{CC},
\end{equation}
where  
\[ V^*_{AA}=v^*(13)+v^*(24)-V_{AA}\]
\[ V^*_{BB}=v^*(14)+v^*(23)-V_{BB}\]
\[ V^*_{CC}=v^*(12)+v^*(34)-V_{CC}.\]
In the following fit $b_0$ is expected to be of the order 
unity and, indeed, fixing $b_0=1$ is found to be a good assumption.

The two-quark potentials $v(ij)$ are taken to be more elaborate than the
three term form of Eq.~\ref{V2}. They are fitted to the lattice data
using \cite{gre:96}
\begin{equation}
\label{Vfit}
v(r_{ij})=0.562 + 0.0696 r_{ij}-  \frac{0.255}{r_{ij}} -
\frac{0.045}{r^2_{ij}}.
\end{equation}
Similarly, the excitation with $E_u$ symmetry is fitted by 
\begin{equation}
\label{V*fit}
\Delta v=v^*(ij)-v(ij) =
\frac{\pi}{r_{ij}} - \frac{4.24}{r^2_{ij}} + \frac{3.983}{r^4_{ij}}.
\end{equation}
The extra terms containing $r^{-2}_{ij}$ and $r^{-4}_{ij}$ are purely for numerical
reasons and ensure that the fitted values of $v(ij)$ and $v^*(ij)$ are,
on average, well within $1 \% $ of the lattice values for all $r_{ij}\ge 2$.

\vspace{0.3cm}   

e) There are two types of off-diagonal element
\begin{equation}
\label{off}
<A^*|V|B^*> \ \ {\rm and} \ \ <A|V|B^*>.
\end{equation}
Unlike $V_{AB}$ in Eq.~\ref{VAB}, there is now no guide from the one-gluon
exchange limit of Eq.~\ref{OGE}. Therefore,  a further assumption is 
necessary for the form of these matrix elements. Several different forms
were tried, but the most successful seems to be the one where the 
matrix elements needed in $[{\bf V}-(E+V_{AA}){\bf N}]=0$
are expressed  as follows

\[<A^*|V+\frac{ f^c}{2}V_{DD}|B^*>=-\frac{ f^c}{2}\left[V_{ABCD}+
c_0V^*_{ABC}\right]=<B^*|V+\frac{ f^c}{2}V_{DD}|A^*>\]
\[<A^*|V+\frac{ f^c}{2}V_{DD}|C^*>=-\frac{ f^c}{2}\left[V_{CABD}+
c_0V^*_{CAB}\right]=<C^*|V+\frac{ f^c}{2}V_{DD}|A^*>
 \]
\begin{equation}
\label{B*C*}
<B^*|V+\frac{ f^c}{2}V_{DD}|C^*>=-\frac{ f^c}{2}\left[V_{BCAD}+
c_0V^*_{BCA}\right]=<C^*|V+\frac{ f^c}{2}V_{DD}|B^*>,
\end{equation} 
where
\[V_{ABCD}=V_{AA}+V_{BB}-V_{CC}-V_{DD} \ , 
\ V_{CABD}=V_{CC}+V_{AA}-V_{BB}-V_{DD},\]
\[V_{BCAD}=V_{BB}+V_{CC}-V_{AA}-V_{DD},\]
\[V^*_{ABC}=V^*_{AA}+V^*_{BB}-V^*_{CC},\]
\[V^*_{CAB}=V^*_{CC}+V^*_{AA}-V^*_{BB} \ , \ V^*_{BCA}=V^*_{BB}+V^*_{CC}-V^*_{AA}.\]
These are seen to be simply the generalisation of Eq.~\ref{VAB} to the
interaction involving excited gluon states.
Also in analogy with $b_0$ in Eqs.~\ref{Vfit}, $c_0$ is a free
parameter -- hopefully of order unity. 

Likewise,
\begin{equation}
\label{A*B}
<A|V+\frac{ f^a}{2}V_{DD}|B^*>=-\frac{f^a}{2}\left[V_{ABCD}+
a_0\frac{V^*_{ABC}}{2}
\right] = <A^*|V+\frac{ f^a}{2}V_{DD}|B> 
\end{equation}
where, again in analogy with $b_0$ in Eqs.~\ref{Vfit}, $a_0$ is a free
parameter -- hopefully of order unity. However, it should be added that
this  hope is even  less well founded than the one for $c_0$.

In the special case of regular tetrahedra
$V_{DD}=V_{AA}=V_{BB}=V_{CC}$ and  ${\bf V}$ reduces to the form
\begin{equation}
\label{KH2}
{\bf V}= \left[ \begin{array}{ccc|ccc}
V_{AA}&-fV_{AA}/2&-fV_{AA}/2&0&-f^aV_a/2&-f^aV_a/2 \\
-fV_{AA}/2&V_{AA}&-fV_{AA}/2&-f^aV_a/2 &0      &-f^aV_a/2     \\
-fV_{AA}/2&-fV_{AA}/2&V_{AA}&-f^aV_a/2 & -f^aV_a/2     &0          \\     \hline
0 &  -f^aV_a/2  & -f^aV_a/2   &V_b&-f^cV_c/2&-f^cV_c/2          \\
-f^aV_a/2 &0    & -f^aV_a/2   &-f^cV_c/2 &V_b     &-f^cV_c/2             \\
-f^aV_a/2 & -f^aV_a/2   &0    &-f^cV_c/2 &   -f^cV_c/2   &V_b       \\ 
\end{array} \right],
\end{equation}

where $V_a=V_{AA}+a_0V^*_{AA}/2 \ , \ V_b=V_{AA}+b_0V^*_{AA} \ , \
V_c=V_{AA}+c_0V^*_{AA}.$ As with all geometries

\begin{equation}
\label{KH1}
{\bf N}= \left[ \begin{array}{ccc|ccc}
1&-f/2&-f/2&0&-f^a/2&-f^a/2 \\
-f/2 &1   &-f/2&-f^a/2 &0     &-f^a/2    \\
 -f/2 & -f/2   &1   &-f^a/2 & -f^a/2     &0        \\     \hline
 0 &-f^a/2    & -f^a/2   &1&-f^c/2&-f^c/2         \\
 -f^a/2 &0    &  -f^a/2  &-f^c/2 &1     &-f^c/2            \\
 -f^a/2 & -f^a/2   &0    &-f^c/2 & -f^c/2     &1         \\      
\end{array} \right].
\end{equation}

The full $6\times 6$ matrix $[{\bf V}-(E+V_{AA}){\bf N}]$ now factorizes into 
three $2\times 2$ matrices,
two of which are identical -- giving the observed degeneracy.
These have the form
\begin{equation}
%\label{KH4}
[{\bf V}-(E+V_{AA}){\bf N}]=\left[ \begin{array}{cc}
-E(1+f/2)&-f^a(E-V_a)/2    \\
 -f^a(E-V_a)/2  &-E(1+f^c/2)+V_b+f^cV_c/2 \\
\end{array} \right]=0, 
\end{equation}
whereas the third $2\times 2$ matrix is
\begin{equation}
%\label{KH4}
[{\bf V}-(E+V_{AA}){\bf N}]= \left[ \begin{array}{cc}
-E(1-f)&f^a(E-V_a)    \\
 f^a(E-V_a)  &-E(1-f^c)+V_b-f^cV_c \\
\end{array} \right]=0. 
\end{equation}
In this case the problem reduces to solving two quadratic
equations for $E$. However, away from the regular tetrahedron the complete 
6$\times$6 matrix needs to be treated directly.
 
\section{Results}
This section is in two parts. In subsection \ref{3.1} the results for
the 6 basis state model at $\beta=2.4$ are given. There it  is shown that
only 5 of the possible 8 parameters have any significant  influence.
Also it is pointed out
that the restriction to the 2 basis state version of subsection \ref{2.1} is
distinctly inferior and that the further restriction to only two-body
interactions (i.e. $k_f=0$ in Eq.~\ref{f}) is for most geometries very
poor. In subsection \ref{3.2}, it is shown that the parameters of the
model do not change significantly as $\beta$ increases i.e. as the
continuum is approached. Therefore, the parameters extracted at
$\beta=2.4$ could be used directly in, for example, a Resonating Group
calculation of a four-quark model of meson-meson scattering as in 
Ref.~\cite{sak:97}.

\subsection{The 6 basis state model at $\beta=2.4$}
\label{3.1}
In Ref. \cite{glpm:96} four quark energies have been extracted for a
variety of geometries using a $16^3\times 32$ lattice with $\beta=2.4$. 
From these energies,  only one hundred -- distributed over six geometries -- are
selected for fitting. Configurations containing flux links of less than two lattice
units were not included, since they have strong lattice artefacts. In 
detail, we use 15 Tetrahedra (T), 6
Squares (S), 12 Rectangles (R), 4 Quadrilaterals (Q), 9 Non-Planar (NP) and 4
Linear (L).
Only the lowest two energies $(E_{0,1})$ from the lattice simulation are used.
But, as mentioned earlier before Eq.~\ref{V2}, the values
of these two energies are more or less the same from the  2$\times$2 and
3$\times$3 simulations. In the latter case a third
energy $(E_2)$ is in fact also available. However,  it is not expected to
be very reliable due to the higher excitations it contains. Its main purpose
was to improve the estimate on $E_{0,1}$ by reducing the contamination from
the higher states. The stability of $(E_{0,1})$ is understandable, since
they are  the lowest states of given symmetry. For example, with squares
their wavefunctions are simply $\psi(E_0)=(A+B)/\sqrt{2}$ and
$\psi(E_1)=(A-B)/\sqrt{2}$.

Before commencing a fit the size of the errors on the above data
must be decided. The lattice simulations, through the boot-strap method,
do indeed produce errors -- statistical ones. However, some estimate
must also be added for systematic errors. How this is done is somewhat
subjective. Here the prescription is to assume all errors must be at
least 0.005 and, also, at least $10\%$ of the eigenvalue itself. 
The former corresponds to about 10\% for the largest values of $E_0$ and
amounts to about 8 MeV.

The above 100 pieces of data are fitted with {\tt Minuit} -- 
the Migrad option. In the  model there are, in principle,
  eight parameters:   $k_f$ in  Eq.~\ref{f}, $f_1^a,f_2^a,k_a$ in 
Eq.~\ref{fa}, $k_c$ in  Eq.~\ref{fc}, $b_0$ in Eqs.~\ref{C*C*} and $a_0, c_0$ 
 in Eqs.~\ref{A*B}.
The strategy for finding the optimal values of these parameters is as
follows.
Firstly, $f^a_1$ is fixed at zero and $b_0$ fixed at unity,
as discussed after Eqs.~\ref{fa} and ~\ref{C*C*}. Furthermore,
preliminary minimizations indicate that $k_c= 0$ is a good approximation
and that $a_0, \ f^a_2$ are strongly correlated.
This latter point  is understandable, since
for regular tetrahedra, where the 6 basis model is needed, the
$V_{ABCD}, \ V_{CABD}$ etc. are zero, so that

\[<A|V+\frac{ f^a}{2}V_{DD}|B^*>=-b_s f^a_2 S \exp(-b_sk_aS)
\frac{a_0V^*_{ABC}}{4}\]
i.e. only the combination $a_0f^a_2$ arises. To avoid this $a_0$ is, at first, 
fixed at four, the value that yields  
essentially the smallest $\chi^2$/dof. This results in the
parameter values shown in the first row of Table~\ref{b24}.
Releasing $k_c$ and $b_0$ to be free parameters gives the results in the
second row. Finally, releasing also $a_0$ gives the third row. 

Table~\ref{b24}  shows the following points:

\vspace{0.2cm}

a) The value of $k_c$ is consistent with zero. This appears to be a
common feature with all fits. This suggests that $f^c\approx 1$, i.e. the
excited configurations interact amongst themselves simply through 2-body
potentials without the 4-quark effect introduced when $k_c\not= 0$. 

b) The second row shows that the expected value of $b_0=1.0$ persists even
when $b_0$ is allowed to vary.
 
c) The third row shows that there is an instability between 
$f^a_2$ and $a_0$, with both having
errors that are $\approx 25\%$ of the parameter value. 
This instability also leads to a value of  $b_0$ that  is much less 
than the expected value of unity.

d) In the first two rows $k_0\approx 3 f^a_2$ -- in qualitative
agreement with the approximate relationship $k_0=4 f^a_2$ derived in the
Appendix.

e) The fourth row comes from Ref.~\cite{gre:97}, which contained a preliminary
fit to the above data, but with a model in which the off-diagonal matrix
elements
of Eqs.~\ref{B*C*}, \ref{A*B} had a slightly different form. The outcome, 
however, is qualitatively rather similar to the present results. 

In order to see from where the $\chi^2$ originates,
Table ~\ref{chi} shows  the contributions of the
average 
\[\chi^2_A=\frac{1}{N(G)}\sum_{i=1}^{N(G)}(E_i-M_i)^2/\Delta E_i\] from
each of the 12 types of data -- i.e. from $E_0,E_1$ for the six
geometries ($T,S,R,Q,NP,L$). 
Here $N(G)$ is
the number of data points for geometry $G$, $M_i$ are the model results and
$E_i\pm \Delta E_i$ are the lattice data. 

Another alternative for $f^a$ -- not supported by the Appendix -- is
\begin{equation}
\label{fa'}
f^a=(f^a_3+f^a_4 \sqrt{S})\exp(-b_sk'_aS). 
\end{equation}
This gives a comparable $\chi^2$/d.o.f. of 1.10, but has the inferior
feature that $f^a_3$ is not consistent with zero and
so in violation of the condition before  Eqs.~\ref{fa}. 

For comparison,  Model Ia with its one free parameter  $k_f$ in Eq.~\ref{f}
gives from {\tt Minuit} a $\chi^2$/d.o.f. of 3.16 for $k_f=0.571(12)$. As seen
from  Table~\ref{chi} most of this $\chi^2$/d.o.f.
comes from the Tetrahedral geometry -- especially from the regular
Tetrahedra. This was already anticipated at the end of  Section~\ref{modII}.
 
Another comparison is the extreme limit of $f=1$, where only two-quark 
potentials are used in Model I. Here all the parameters are fixed, so that
this is not a minimization.
From Table~\ref{chi} it is seen that
this gives a very poor representation of the data -- especially for
$E_1$. As expected, the larger the size of the geometry the larger the
contribution to $\chi^2$/d.o.f., since those are the cases that need most
the smaller values of $f$. For example, with a 6$\times$6 square the
model  gives $E_{0,1}=-0.124,0.3721$, whereas from the lattice
simulation the corresponding energies are $-0.028(5),0.065(7)$.  
This comparison is the basis of our statement
in the introduction that multi-quark models containing only the standard
two-quark potential of Eq.~\ref{V2} or even the more elaborate form in
Eq.~\ref{Vfit} seem to have no justification for the cases considered here.

\subsection{The continuum limit}
\label{3.2}
The above results are for $\beta =2.4$, which corresponds to a lattice
spacing of $a\approx 0.12$ fm. But, in practice, if the model is to
be used, for example, in some type of resonating group approach
\cite{sak:97}, then it is the continuum limit that is needed. Such an
extrapolation would, in principle, require repeating the above lattice 
calculations with all 6 geometries for a series of increasing values of 
$\beta$. This we do not plan to do here. Instead we exploit the
conclusions of Ref.~\cite{pen:96b}. There it was shown that the four-quark
{\em binding energies} already scale at $\beta=2.4$, eventhough the total
two- and four-quark energies do not scale separately. Therefore, here we
simply assume that, if some binding energy $E(\beta)$ 
 would be  scaled to the $\beta=2.4$ lattice by the transformation
$\rho E[\beta,r(\beta)/\rho ]\rightarrow E(2.4), \ \ {\rm where } \ \
\rho=a(2.4)/a(\beta)$, then we would find that this scaled $E(2.4)$ equals the
binding energy $E(r)$ we have  calculated directly  on the $\beta=2.4$
lattice. Here the $r(\beta)$ and $r$ are distances in the lattice units of
the two different lattices.
However, to completely specify the model at some other value of $\beta$ also needs 
a knowledge of $v(ij,\beta)$ and $\Delta v(ij,\beta)$ corresponding to the
$\beta=2.4$ expressions in Eqs.~\ref{Vfit}, \ref{V*fit}. In Ref.~\cite{pen:96b} it
was shown that $v(ij,\beta)$ did not scale. But this is not a problem,
since $v(ij,\beta)$ is documented in the literature for
other values of $\beta$ -- see Refs.~\cite{pen:97b,bal:94,boo:92}. These 
can then be scaled in the same way as
the $E(\beta)$ to give $v(ij,2.4)$. However, as found in Ref.~\cite{pen:96b}
we do not expect these $v(ij,2.4)$ to equal  the $v(ij)$ in Eq.~\ref{Vfit}. 
It remains to specify $\Delta v(ij,\beta)$.
Here we simply assume that it scales, so that the values at $\beta=2.4$
can be  used directly in precisely the same way as the binding energies
$E(\beta)$.
The rational behind this last assumption is twofold:

%i) The 6 basis model involving $\Delta v$ is an extension of the
%original 2 basis model, which is itself qualitatively correct for most 
%geometries. Only for the tetrahedron geometry does it give a qualitative
%difference. The  $\Delta v$ terms  are, therefore, essentially
%corrections  -- albeit important in a few  cases -- to the original 2
%basis model.    

i) In the model it is the  {\em differences} $\Delta v$ and 
$v(ij,\beta)-v(kl,\beta)$ that
enter as in Eqs.~\ref{C*C*}, \ref{B*C*} and \ref{A*B}. Therefore, some of the
lack of 
scaling in $v^*(ij,\beta)$ and $v(ij,\beta)$ may well cancel -- as in the
four-quark binding energies.

ii) Any lack of scaling that remains  will be reflected in a lack of scaling
of the $a_0,b_0,c_0$ multiplying the $\Delta v$ and  $V^*$'s.

The actual values of $\beta$ used are 2.5\cite{pen:97b},
2.5 (denoted by $2.5'$), 2.635 and 2.74\cite{bal:94} and 2.85\cite{boo:92}. 
In the last case, the authors
present two potentials[$V(R,5)$ and $V(\infty)$] that are appropriate for
interquark distances $R$ greater than $6a(2.85)\approx 0.17$ fm. Their own
preference is  $V(R,5)$ -- the one we use.

The results are shown in Table~\ref{Minuit4}. The
main observation is that, of the free parameters, only
$k_f$  appears to be a significant function of $\beta$ 
 -- but even there all values are  within the error bars.
Therefore, the final version of the model has only four independent parameters.

A similar continuum extrapolation was carried out in
Ref.~\cite{pen:96b}.
However there are several major differences compared with the above
analysis:

a) Much less data for each value of $\beta$ was included in the fits, since 
only Squares and Tilted Rectangles were analysed. Furthermore,
only the $E_0$'s  were fitted. This means that, for each $\beta$, there
were only 17 pieces of data compared with the present case of 100 pieces
covering 6 geometries with both $E_0$ and $E_1$. 

b) All the two- and four-quark energies were simulated at five values
of $\beta$ and interpolated to the same physical sizes. As scaling was 
found for the binding energies, they are here 
obtained directly from the $\beta=2.4$ values. 

c) In order to study scaling and the effect of lattice artefacts, the errors 
on the data were taken to be purely statistical and so
were much smaller than the present ones.
 
d) The model analysis was carried out with the $2\times 2$ model. Since
this results in a quadratic equation for the energies, the step
$E_0\rightarrow f$ could be done analytically. After this the parameters
$k_f,\ k_P,\ f_0$ in Eq.~\ref{fIb}  were extracted by a fitting procedure.

\subsection{A comparison with other work}
\label{3.3}

A fit with a potential model to some of the above
four-quark binding energies  was attempted earlier  in  Ref.~\cite{fur:95}.
However, there are several important differences compared with the 
present fit:

1) The model contained only $2\times 2$ or $3\times 3$ bases that were
essentially constructed from $A,B,C$.

2) The Tetrahedral data was not included.

3) Most of the data was for $\beta=2.4$. Only squares included that from
$\beta=2.5.$

4) The overlap factor is that in Eq.~\ref{fIb} but with $f_0=1$.

 For the four geometries S,R,Q,L the area $S$ is the same as
that used here in Eq.~\ref{Def:S} and for geometry NP a curved surface
is used. However, the perimeters $P$ are determined by the underlying
lattice structure of the geometry and not the usually expected
perimeter. For example, for Quadrilaterals, $P$ is the perimeter of the
{\em rectangle} enclosing the quadrilateral. The reason for this choice is an
attempt to take into account lattice artefacts, which are present for
the smallest configurations. These artefacts are more important in 
Ref.~\cite{fur:95}, since configurations containing flux-lines of only one
lattice link are included. In the present work the cut-off is at two
lattice links. For Squares the outcome in ~\cite{fur:95} was $k_f=0.296(11)$ 
and
$k_P=0.080(2)$, compared with $k_f=0.571(12)$ using Model Ia above in
Subsection~\ref{2.1}. The difference between the values of these area 
constants is because the perimeter term, which is introduced to make
artefact corrections for  {\em small} configurations, is still important for
large configurations, where rotational symmetry is restored and their
presence not needed. For example, with a square of side $d=5$, the two
terms in the exponent of Eq.~\ref{fIb} are comparable, even though the
lattice data for this square is --0.042(5) and --0.047(5) for the
corresponding tilted square. In the lattice simulation the basis states
have a similar form compared with the ones used in ~\cite{fur:95} but with
the major difference that they are fuzzed to such an extent that these
underlying basis states get transformed. Therefore, in ~\cite{fur:95}, for
larger configurations it would be necessary to include many other terms
with different perimeters, so that on the average the usual perimeter
for a particular geometry would probably be more appropriate. This latter 
choice
was tested in Ref.~\cite{pen:96b}, where it was found to reduce in the
continuum limit to the form of Model Ia. Thus the perimeter term seems to 
measure lattice artefacts also when the physical perimeter is used.
 
In Ref.~\cite{fur:96}, the model in Ref.~\cite{fur:95} is extended to include 
the
tetrahedral data by using an $8\times 8$ basis, where now all the states are
constructed in terms of the underlying lattice. This, therefore, suffers
from the above problem that the results are always dependent on the
lattice even for large configurations, where rotational invariance is restored.

\section{Conclusions}
In this article an interquark potential model is constructed than can
explain, on the average, the energies of a series of four-quark systems,
in which the four static quarks are in any one of six geometries. The
input energies used as data for this construction \cite{glpm:96}
were calculated on a $16^3\times 32$ lattice for four static SU(2) quarks
with $\beta=2.4$. Two or three basis states were used in the lattice
simulation enabling reliable estimates to be made of the lowest two
energy eigenvalues ($E_{0,1}$). In all, 100 pieces of data were considered
suitable for confronting the model. Other pieces were rejected, if they
involved configurations suspected of having strong lattice artefacts (e.g. with
states constructed from less than two lattice links) or had vanishingly
small binding energies (e.g. elongated rectangles with $r\gg d$).

The full model utilized 6 basis states $A,B,C,A^*,B^*,C^*$ and 
in its most general  form  had eight parameters. 
However, in practice, only 4 of these ($k_f,k_a,f^a_2$ and $c_0$) need be
considered as free when fitting the data.

The parameters that are, perhaps, of most interest are those connected
with the ranges of the various interactions, namely, $k_f,k_a, k_c$.
Here we define 'range' as $r_{f,a,c}=\sqrt{1/b_sk_{f,a,c}}$. 
In Model Ia, where $k_{a,c}$ are effectively infinite, we get
$k_f(Ia)$=0.57(1) i.e. $r_f(Ia)=5.0$ in lattice units.
However, in Model III as the excited states $A^*,B^*,C^*$ are introduced, the
interaction between the basic states $A,B,C$ decreases by raising
$k_f$ to  1.51 giving $r_f(III)=3.1$. 
But at the same time this loss of binding by the
direct interaction between $A,B,C$ is compensated by their coupling to the
$A^*,B^*,C^*$ states. This coupling is found to have about the 
{\em same}  range $r_a=5.1$ as $r_f(Ia)$ above, whereas the direct
interaction between the $A^*,B^*,C^*$ states seems, in all fits, to be 
satisfied with simply a two-quark description without any four-quark 
correction (i.e. $k_c$=0). The observation that $r_f(Ia)\approx r_a$
suggests that the energy density has a range dictated by the longest
range available -- namely $r_a$. Therefore, when the 
$A^*,B^*,C^*$ states are not explicitly present, as in Model Ia, the
only available range $r_f(Ia)$ has to simulate the role of $r_a$.
In the binding energies the contributions from the 
$A^*,B^*,C^*$ states rapidly dominate over those from the 
$A,B,C$ states. For example, with squares of side $R$, the $A,B,C$
states contribute only 85, 40, 10\% to the binding energy for $R$=2,4,6
respectively. Of course, at the largest distances ($\approx 0.7$ fm)
the quenched approximation is expected to break down and the role
of quark-pair creation to become  important. 
We, therefore, come to the following scenario for the four-quark
interaction.
At the shortest distances, upto about 0.2 fm, perturbation theory is
reasonable with the binding being given mainly by the $A,B,C$ states
interacting simply through the two-quark potentials with little
effect from four-quark potentials. However, for intermediate ranges, from
about 0.2 to 0.5 fm, the four-quark potentials act in such a way as to
reduce the effect of the $A,B,C$ states so that the binding is dominated
by the $A^*,B^*,C^*$ states, which now  interact amongst themselves again 
simply through the two-quark potentials {\em with little
effect from four-quark potentials}. At larger ranges quark-pair creation
can no longer be neglected.

The next step in this work would be to carry out a Resonating Group
calculation in the spirit of Ref.~\cite{sak:97}. This would then test the
universality of the above model by extending it from the 6 geometries
discussed here to general four-quark geometries.

\section{Acknowledgement}
The authors wish to thank Drs. S. Furui and B. Masud for useful discussions.	
Funding from the Finnish Academy and Magnus Ehrnrooth foundation (P.P.) is gratefully 
acknowledged. Our simulations were performed on the Cray C94 at the Center 
for Scientific Computing in Espoo.

\vspace{1.2cm}

{\Large {\bf Appendix}}

\vspace{0.7cm}

{\bf The justification of Eq.~\ref{fa}}

\vspace{0.3cm}

In the case of regular tetrahedra a simple model for the 
$A(13,24),A^*(13,24)$ states is as follows:

\vspace{0.1cm}

\[A(13,24)=N\frac{1}{\sqrt{2}}(ab+cd)\frac{1}{\sqrt{2}}(ef+gh),\]

\vspace{0.1cm}

\[B(14,23)=N\frac{1}{\sqrt{2}}(aj+ih)\frac{1}{\sqrt{2}}(ek+ld),\]

\vspace{0.1cm}

\[A^*(13,24)=N^*\frac{1}{\sqrt{2}}(ab-cd)\frac{1}{\sqrt{2}}(ef-gh),\]

\vspace{0.1cm}

\begin{equation}
\label{ABdef}
B^*(14,23)=N^*\frac{1}{\sqrt{2}}(aj-ih)\frac{1}{\sqrt{2}}(ek-ld),
\end{equation}   

\vspace{0.1cm}

where  $a,b,c...k,l$ are links along the  sides of the cube defining the regular
tetrahedron and $N$ and $N^*$ are normalisation constants. The loops
$a,b,c,d$ and $e,f,g,h$ are taken to be in the  $xz$ plane and the loops
$a,j,i,h$ and $e,k,l,d$ are in the $yz$ plane.

The above definitions guarantee $<A(13,24)A^*(13,24)>=0$ etc.
and give for the two normalisations: 

$<AA>=N^2[1+1+S(xz)+S(xz)]^2/4$ i.e. $N=1/[1+S(xz)]$ and

\vspace{0.1cm}

$<A^*A^*>=N^{*2}[1+1-S(xz)-S(xz)]^2$ i.e. $N^*=1/[1-S(xz)]$,

\vspace{0.1cm}

where $S(xz)$ is a $R\times R$ Square loop of links in the $xz$ plane along
the sides of the cube i.e.
perimeter is of length 4$R$. Other loops needed in this discussion are
as follows:

The loop $L$ is two connected faces of the cube in an $L$-shape i.e. perimeter
is of length 6$R$, 

$C$ is three connected faces of the cube giving a corner i.e. perimeter
is also of length 6$R$

$T$  is three faces of the cube giving a "table" i.e. perimeter
is of length 8$R$

\vspace{0.1cm}

With this notation the following overlaps can be written down:

\vspace{0.1cm}
 $<A(13,24)B(14,23)>=-\frac{1}{2}\frac{1}{4}[2S+2T+4C+8L]/(1+S)^2=-f/2$,

\vspace{0.1cm}
where the initial factor of --1/2 is the usual colour factor  giving

\vspace{0.1cm}

\[f=\frac{[2S+2T+4C+8L]}{4(1+S)^2}.\]

Likewise,
\vspace{0.1cm}

 \[f^c=\frac{[2S+2T+4C-8L]}{4(1-S)^2} \ \ {\rm and} \ \
f^a=\frac{[2S+2T-4C]}{4(1-S^2)}.\]

\vspace{0.1cm}

A rough estimate of the relative importance of $S,L,C,T$ can be seen as
follows:
In a Wilson  loop, the further two combinations of links (e.g.$L1,L2$)
are
separated in Euclidean time, the less is their correlation
$<L1.L2>=\exp(-VT)$.
Similarly, spatial links are less correlated the further they are apart.

\vspace{0.1cm}

For the Square ($S$)  let side 1 have a factor 1 and the two adjacent
sides factors of
$p$ and the fourth side $p^2$. This makes $S=D.1.p.p.p^2=p^4$, where $D$
is
some constant and $p$ some correlation factor to adjacent sides with
$p<1$

\vspace{0.1cm}

For the $L$-shape let the correlation factors have an extra  $q$, when
changing
direction out of the original side of the cube and an extra $r$ for the final
side.
In all $L=D.1.p.pq.pqr.pq.p=Dp^5q^3r$ when moving around the perimeter
of the $L$-shape.

\vspace{0.1cm}

For the $C$-shape the same factor arises as in the $L$-shape.
 This suggests $L=C$.

\vspace{0.1cm}

For the $T$-shape the last set of links is close to original ones, so
that
the factor on the last link is $p^2$ as in the $S$ case. In all
$T=D.1.p.pq.pqs.p^2.pqs.pq.p=Dp^8q^4s^2$.

\vspace{0.1cm}

For large $R$ we expect $p,q,r,s\ll 1$ which suggests $S\gg (L \ or \ C)
\gg T$.

For small $R$ we expect $p,q,r,s\approx 1$ i.e. $S\approx (L \ or \ C)
\approx T \approx 1$

One way to parametrize these two limits is to take

\vspace{0.1cm}

$S=\exp(-WR^2)\approx (1-WR^2)$ since it involves a single square of area
$R^2$.

\vspace{0.2cm}

$L=\exp(-W2R^2)\approx (1-2WR^2)$ since it involves two squares of area
$R^2$.

\vspace{0.2cm}

$T=\exp(-W3R^2)\approx (1-3WR^2)$ since it involves 3 such squares.

\vspace{0.2cm}
 
Keeping only the terms upto  $O(R^2)$ in $S,L,C,T$ gives simply $f=1-WR^2$,
$f^a=f^c=0$ and is model independent. Whereas the addition of the
$O(R^4)$ terms from expanding further the exponentials gives $f^a=WR^2/4$ and
$f^c=1/2$. This last result  
clearly shows that the $O(R^4)$ terms are not so
simple, since we expect $f^c\rightarrow 1$ as $R\rightarrow 0$ -- as in the case
of $f$. Similarly these arguments would suggest $k_f=4f^a_2$.
However, the $R^2$ dependence of $f^a$ is probably correct and
so for general geometries we assume $f^a\propto {\rm Area}$ as 
${\rm Area}\rightarrow 0$.
For the case of regular tetrahedra,  the appropriate area from
Eq.~\ref{Def:S} is $\sqrt{3}R^2$. 

\newcommand{\href}[2]{#2}\begingroup\raggedright\endgroup

\pagebreak

\mediumtext  
\begin{table}[htb]
\begin{center}
\begin{tabular}{ccccccccc}
&$k_f$   &$k_a$   &$f_2^a$   &$k_c$   &$a_0$   &$b_0$    &$c_0$&$\chi^2$/dof\\ \hline
Fix 4
&1.51(8)&0.55(3)&0.51(2)&\underline{0.0}&\underline{4.0}&\underline{1.0}
&3.9(3)&1.03\\
Fix 2
&1.50(8)&0.57(3)&0.54(7)&0.02(5)&\underline{4.0}&1.06(32)&4.3(1.3)&1.05\\
Fix 1   &1.61(11)&0.59(5) &0.28(6)   &0.00(11)&3.4(8)  
&0.19(2)  &0.78(6)&0.90 \\
Ref.~\cite{gre:97} &1.25(6)&0.54(11)&0.66(4)&0.04(20)&4.4(3)&2.2(6)
&8.0(4)&1.08\\
\end{tabular}
\caption{ The values of the parameters defining the interaction and
fixing $f_1^a=0$. Those numbers underlined are kept fixed. }
\label{b24}
\end{center}
\end{table}

\mediumtext
\begin{table}[htb]
\begin{center}
\begin{tabular}{ccccccc}
Model&\multicolumn{2}{c}{Model III}&\multicolumn{2}{c}{Model Ia}
&\multicolumn{2}{c}{Model I(f=1)}\\ \hline
Geometry& $\chi^2_A(E_0)$&$\chi^2_A(E_1)$& $\chi^2_A(E_0)$
&$\chi^2_A(E_1)$
& $\chi^2_A(E_0)$&$\chi^2_A(E_1)$\\ \hline 
T&0.9&0.7&8.5&7.6&15.8&141.5\\
S&0.2&1.9&1.2&1.5&102.4&659.6\\
R&1.1&0.9&1.1&0.5&79.2&527.2\\
Q&2.7&1.2&1.0&0.3&0.8&6.2\\
NP&1.0&0.2&1.7&0.2&0.6&16.9\\
L&0.3&2.6&0.3&2.6&0.3&2.6 \\ 
\end{tabular}
\caption{The contributions $\chi^2_A(E_{0,1})$ from the two states ($E_0,E_1$)
in each
of the six geometries ($T,S,R,Q,NP,L$) to the total $\chi^2$/d.o.f. of 1.03
for Model III, 3.16 for Model Ia and 144 for Model I($f=1$).}

\label{chi}
\end{center}
\end{table}

\mediumtext
\begin{table}[htb]
\begin{center}
\begin{tabular}{ccccccc}
Parameter& $\beta=2.4$&$\beta=2.5$&$\beta=2.5'$&$\beta=2.635$&
$\beta=2.74$&$\beta=2.85$ \\ \hline
$k_f$&1.51(8)&1.48(8)&1.50(8)&1.49(8)&1.45(8)&1.43(8)\\
$k_a$&0.55(3)&0.56(3)&0.56(3)&0.56(3)&0.57(3)&0.57(3)\\
$f^a_2$&0.51(2)&0.51(2)&0.51(2)&0.51(2)&0.51(2)&0.51(2)\\
$c_0$&3.9(3)&3.9(3)&3.9(3)&3.9(3)&3.9(3)&3.9(3)\\ \hline
$\chi^2$/d.o.f.&1.03&1.03& 1.11&1.09&1.15&1.07 \\ 
\end{tabular}
\caption{The values of the parameters defining the interaction fixing
$f_1^a=k_c=0, \  b_0=1.0$ and $a_0=4.0$ for a series of $\beta$'s.}
\label{Minuit4}
\end{center}
\end{table}

\end{document}